# Kinetics of precessing ball solitons in ferromagnet at the first-order transition

**V.V.Nietz**

*Frank and Shapiro Laboratory of Neutron Physics, Joint Institute for Nuclear Research, Dubna, Moscow region 141980, Russia*

## Abstract

The fundamentals of precessing ball solitons (PBS) arising as a result of the energy fluctuations at the first-order phase transition induced by a magnetic field in ferromagnets with uniaxial anisotropy are presented. When external magnetic field is anti-parallel to the magnetization direction of the crystal, PBS states are possible in a wide range of amplitudes and energies, including the positive and negative energy relative to an initial condition. PBS are born with the greatest probability at near-zero energy, i.e. near the bifurcation point. Evolution of the PBS, at which they transform into macroscopic domains of a new magnetic phase or into quasi-equilibrium solitonic state, is analyzed.

*PACS*: 75.25; 75.50.E; 29.30.H

*Keywords*: Ferromagnet, Single crystal, Magnetic field, First-order phase transition, Kinetics of the phase transition, Precessing ball soliton, Evolution of soliton.

Tel: +7-496 21-65-552
Fax: +7-496-21-65-882
*E-mail address*: nietz@nf.jinr.ru

## Introduction

Several articles [1-6] were devoted to spherical magnetic solitons in the first-order phase transitions. The precessing magnetic solitons (precessing ball solitons, PBS) in magnetic crystals with the anisotropy of the type “easy axis” or “easy plane” represent a particular interest. Presence of only one axis of anisotropy leads to precession of magnetic moments relatively to this axis and, as a consequence, to magnetic solitons with a continuous spectrum. In such cases, the precession of moments has a stabilizing effect on the existence of three-dimensional solitons with various amplitudes and energy values, respectively. In oriented phase transitions induced by an external magnetic field, the energy of PBS with a ellipsoid-of-rotation symmetry decreases abnormally near the bifurcation points, and, hence, such solitons can arise spontaneously as result of thermal fluctuations. PBS in the first-order phase transitions can become germs of a new phase. Under certain conditions they grow and turn into macroscopic domains of a new phase.

Precessing solitons in the phase transitions induced by a field have been considered earlier in two processes: the spin-flop transition in an antiferromagnet [1, 3–6] and the phase transition in ferromagnet with a “light plane” anisotropy [2]. In the given article, the PBS at phase transition in ferromagnet of the “light axis” type, when the external field is directed along the axis of anisotropy but opposite to the direction of initial magnetization, are analyzed.

## 1. Equations for PBS

To analyze magnetic solitons in the ferromagnet with uni-axial anisotropy, we use the following equation and corresponding expression for the thermodynamic potential:

$$\frac{1}{\gamma}\frac{\partial \mathbf{m}}{\partial t} = -\mathbf{m}\times\tilde{\mathbf{H}} + \frac{\kappa}{\gamma}\left(\mathbf{m}\times\frac{\partial \mathbf{m}}{\partial t}\right) \tag{1}$$

$$W = \frac{K_1}{2}|m_\perp|^2 - m_z H_z + \frac{\alpha}{2}\left[\left(\frac{\partial \mathbf{m}}{\partial X}\right)^2 + \left(\frac{\partial \mathbf{m}}{\partial Y}\right)^2\right] + \frac{\alpha_z}{2}\left(\frac{\partial \mathbf{m}}{\partial X}\right)^2 \tag{2}$$

Here $H_z$ is an external magnetic field directed along the anisotropy axis $Z$ ($H_z > 0$). In this paper, we will consider the situation when it is possible to use the approximation of homogeneity of this field. $\mathbf{m}$ is a non-dimensional vector of ferromagnetism equal in the absolute value to *1*; $m_\perp = m_x + i m_y$; $K_1 > 0$, $\kappa > 0$, $\gamma = \frac{2\mu_B}{\hbar}$;

$$\tilde{\mathbf{H}} = -\frac{\partial W}{\partial \mathbf{m}} + \frac{\partial}{\partial X}\frac{\partial W}{\partial\left(\frac{\partial \mathbf{m}}{\partial X}\right)} + \frac{\partial}{\partial Y}\frac{\partial W}{\partial\left(\frac{\partial \mathbf{m}}{\partial Y}\right)} + \frac{\partial}{\partial Z}\frac{\partial W}{\partial\left(\frac{\partial \mathbf{m}}{\partial Z}\right)}.$$

Equation (1) can be written as follows:

$$i\frac{\partial m_\perp}{\partial \tau} = -h m_\perp - m_z m_\perp + m_z \Delta m_\perp - m_\perp \Delta m_z + \kappa\left(m_\perp \frac{\partial m_z}{\partial \tau} - m_z \frac{\partial m_\perp}{\partial \tau}\right) \quad (3)$$

In Eq. (3), the differentiation is carried out with respect to the dimensionless time $\tau = 2\mu_B K_1 \hbar^{-1} t$ and dimensionless coordinates $x = K_1^{0.5}\alpha^{-0.5}X$, $y = K_1^{0.5}\alpha^{-0.5}Y$, $z = K_1^{0.5}\alpha_z^{-0.5}Z$, $h = H_z / K_1$. In our case $m_z = \pm\sqrt{1-|m_\perp|^2}$, such normalization permits to limit oneself to the scalar (3) equation only.

Multiplying Eq. (3) by $\frac{\partial m_\perp^*}{\partial \tau}$ value, producing the complex conjugate of the equation obtained, and then summing up the two equations, we obtain the relation:

$$\frac{d}{d\tau}\left[\frac{1}{2}|m_\perp|^2 \pm h\sqrt{1-|m_\perp|^2} + \frac{1}{2}|\nabla m_\perp|^2 + \frac{1}{2}\left|\nabla\sqrt{1-|m_\perp|^2}\right|^2\right] =$$
$$= \nabla\left[\left(\frac{\partial m_\perp}{\partial \tau}\nabla m_\perp^* + \frac{\partial m_\perp^*}{\partial \tau}\nabla m_\perp\right) + \left(\nabla\sqrt{1-|m_\perp|^2}\,\frac{\partial}{\partial \tau}\sqrt{1-|m_\perp|^2}\right)\right] - \kappa\left(\left|\frac{\partial m_\perp}{\partial \tau}\right|^2 + \left|\frac{\partial}{\partial \tau}\sqrt{1-|m_\perp|^2}\right|^2\right) \quad (4)$$

Here the first sign from two is used for the case $m_z = -\sqrt{1-|m_\perp|^2}$, but the second one is for $m_z = +\sqrt{1-|m_\perp|^2}$ case. Hence, the invariant of energy density for Eqs. (1) and (3) is the following:

$$e(\mathbf{r},\tau) = \frac{1}{2}|m_\perp|^2 \pm h\sqrt{1-|m_\perp|^2} + \frac{1}{2}|\nabla m_\perp|^2 + \frac{1}{2}\left(\nabla\sqrt{1-|m_\perp|^2}\right)^2, \quad (5)$$

and the change of energy density due the dissipation is the following:

$$\left(\frac{de(\mathbf{r},\tau)}{d\tau}\right)_{diss} = -\kappa\left(\left|\frac{\partial m_\perp}{\partial \tau}\right|^2 + \left|\frac{\partial}{\partial \tau}\sqrt{1-|m_\perp|^2}\right|^2\right). \quad (6)$$

To obtain another invariant, we multiply Eq. (3) by $m_\perp^*$, make the complex conjugate of the equation obtained, and then take the difference between the two equations. As a result, the following relation can be received:

Formatted: Font: Bold

$$\pm\frac{d}{d\tau}\sqrt{1-|m_\perp|^2} = \frac{1}{2i}\nabla\left(m_\perp^*\nabla m_\perp - m_\perp\nabla m_\perp^*\right) + \frac{\kappa}{2i}\left(m_\perp\frac{\partial m_\perp^*}{\partial \tau} - m_\perp^*\frac{\partial m_\perp}{\partial \tau}\right). \quad (7)$$

Hence, we have the invariant of effective density of magnetic moment corresponding to Eq. (3):

$$m_z(\mathbf{r},\tau) = \pm\sqrt{1-|m_\perp|^2}\ ; \tag{8}$$

and the expression for the change of this moment due to dissipation is:

$$\left(\frac{dm_z(\mathbf{r},\tau)}{d\tau}\right)_{diss} = \frac{\kappa}{2i}\left(m_\perp \frac{\partial m_\perp^*}{\partial \tau} - m_\perp^* \frac{\partial m_\perp}{\partial \tau}\right). \tag{9}$$

The solutions of Eq. (3) have the following form:

$$m_\perp(\mathbf{r},\tau) = p(\mathbf{r},\tau)e^{i(k(\tau)x-\omega(\tau)\tau)}, \tag{10}$$

it is supposed that the excitations advance along the $x$-axis. Therefore, from Eq. (3) we have the following two equations:

$$\Delta p + \frac{p}{1-p^2}(\nabla p)^2 = p\sqrt{1-p^2}\left[\left(1+k^2\right)\sqrt{1-p^2} \pm \left(h + (\omega-\Delta)\right)\right] + \kappa\frac{\partial p}{\partial \tau}, \tag{11}$$

$$\frac{\partial p}{\partial \tau} = \pm\sqrt{1-p^2}\left(\kappa(\omega-\Delta)p - 2k\frac{\partial p}{\partial x}\right), \tag{12}$$

where $\Delta = \left(\frac{dk}{d\tau}x - \frac{d\omega}{d\tau}\tau\right)$. In Eqs. (11) and (12), the (+) sign is used for $m_{zm} > 0$ case; but the (-) sign, for $m_{zm} < 0$ case.

The change of energy connected with dissipation is

$$\left(\frac{de(\mathbf{r},\tau)}{d\tau}\right)_{diss} = -\kappa\left[\left(2k\frac{\partial p}{\partial x} + \kappa p\omega\right)^2 + \omega^2 p^2\left(1-\kappa^2 p^2\right)\right] \cong -\kappa\left[4k^2\left(\frac{\partial p}{\partial x}\right)^2 + \omega^2 p^2\right]. \tag{13}$$

Equation (11) can be transformed to equation for $m_z$:

$$\Delta m_z + \frac{m_z}{1-m_z^2}(\nabla m_z)^2 = -\left(1-m_z^2\right)\left[\left(1+k^2\right)m_z + \left(h + (\omega-\Delta)\right)\right] + \kappa\frac{\partial m_z}{\partial \tau}, \tag{14}$$

At $k \neq 0$ the systems of Eqs. (11)–(14) correspond to localized but non-spherical excitations. Only at $k \equiv 0$, PBS are of spherical symmetry. Equations for such PBS are the following:

$$\frac{d^2 m_z}{dr^2} + \frac{2}{r}\frac{dm_z}{dr} + \frac{m_z}{1-m_z^2}\left(\frac{dm_z}{dr}\right)^2 \cong -\left(1-m_z^2\right)\left[m_z + \left(h + \omega(\tau)\right)\right], \tag{15}$$

$$\left(\frac{de(r,\tau)}{d\tau}\right)_{diss.\omega} \cong -\kappa\omega(\tau)^2\left(1-m_z^2\right), \tag{16}$$

$$\frac{\partial m_z}{\partial \tau} = -\kappa\left(1-m_z^2\right)\omega(\tau). \tag{17}$$

In Eqs. (15) - (17), the replacement $\left(\omega + \frac{d\omega}{d\tau}\tau\right) \rightarrow \omega(\tau)$ has been made. It is admissible, if to consider a rather slow change of the precession frequency and limit oneself to the linear term in the expansion over time. Furthermore, in Eq. (15) we take in account that $\kappa^2 << 1$ (in the following examples, we use the value of the dissipation parameter $\kappa = 0.0005$ proceeding from the typical experimental data on ferromagnetic resonance (see, for example, in [7])).

## 2. PBS taking into account the demagnetizing fields

The effective field includes external magnetic field $H_z$, the demagnetizing field $H_{dem0}$ connected with a form of crystalline sample, and the demagnetizing field $H_{dems}$ created by self-soliton. Here we suppose that $H_{dem0}$ field is homogeneous in the PBS volume. Correspondingly, if the sample is an ellipsoid of rotation and its axis is directed along the axis of symmetry: $H_{dem0} = N_0 M_0$, where $N_0$ is the factor of demagnetization for the sample, $\mathbf{M}_0$ is magnetization of a crystal. The problem becomes complicated with $H_{dems}$ because of the magnetization in soliton volume is not homogeneous. Only in the limited case of PBS with amplitude of $m_{zm} \cong +1$, when radius of soliton is considerably larger than the thickness of its boundary layer, it is possible to use the approximation of uniform distribution of magnetization in PBS volume. In such utmost case, when PBS have the ellipsoid-of-rotation symmetry, the PBS have own demagnetizing field equal to $H_{dems} \cong -N_s |\mathbf{M}_s|$. In the given paper, to consider the influence of own demagnetizing field, taking into account that PBS is a part of the magnetized crystal, we use the approximate expression: $H_{dems} \cong -N_s \left( |\mathbf{M}_0| + \mathbf{M}_{sz} \right)$. In such case, the energy connected with own demagnetizing field of PBS is

$$E_{dem,s} = \frac{1}{2} N_s (1 + m_z)^2 M_o^2 . \tag{18}$$

Let us assume that the sample is used in the form of a flat (plane) plate perpendicular to the anisotropy axis, and PBS are of a spherical symmetry. In this case, in Eq. (2), instead of $(-m_z H_z)$ we should use the following addition:

$$\Delta W = -m_z (H_z + 4\pi M_0) + \frac{1}{2} \frac{4\pi}{3} (1 + m_z)^2 M_0 . \tag{19}$$

Note that the (18) expression for own demagnetization energy of PBS is correct also in the other limiting case, namely for PBS with very small amplitudes, when at $q_m \to 0$ the radius of PBS increases, i.e. $r_{1/2} \to \infty$.

As a result, we obtain the equation for PBS in the following form:

$$\frac{d^2 m_z}{dr^2} + \frac{2}{r}\frac{dm_z}{dr} + \frac{m_z}{1-m_z^2}\left(\frac{dm_z}{dr}\right)^2 \cong -\left(1-m_z^2\right)\left[(1-D)m_z + \left(h - D + \omega(\tau)\right)\right]. \tag{20}$$

In this equation $h = \left(h_z + 4\pi M_0 / K_1\right)$, $D = \frac{4\pi}{3}\frac{M_0}{K_1}$. In such case, the energy of PBS equals

$$E_s = P\int e(r) r^2 dr = P\int\left[\frac{1-m_z^2}{2} - (1+m_z)h + \frac{D}{2}(1+m_z)^2 + \frac{1}{2\left(1-m_z^2\right)}\left(\frac{dm_z}{dr}\right)^2\right] r^2 dr \tag{21}$$

where $P = 4\pi M_0 K_1^{-0.5}\alpha^{1.5}$.

## 3. PBS characteristics

Here we consider PBS without the dissipation, i.e. for $\kappa = 0$.

We discuss the solutions of Eq. (20) at a phase transition, i.e. when initially the ferromagnet was magnetized along the $(-z)$ direction ($m_{z0} = -1$), but a field is directed along the $(+z)$ direction. In Fig.1, the forms of PBS are shown at different frequency for $h = 0.998$. In this figure the soliton number 3 corresponds to the maximum energy in the spectrum of PBS (except the values at $\omega \to (1-h)$. Energy of the soliton number 4 equals about zero and corresponds to the bifurcation point.

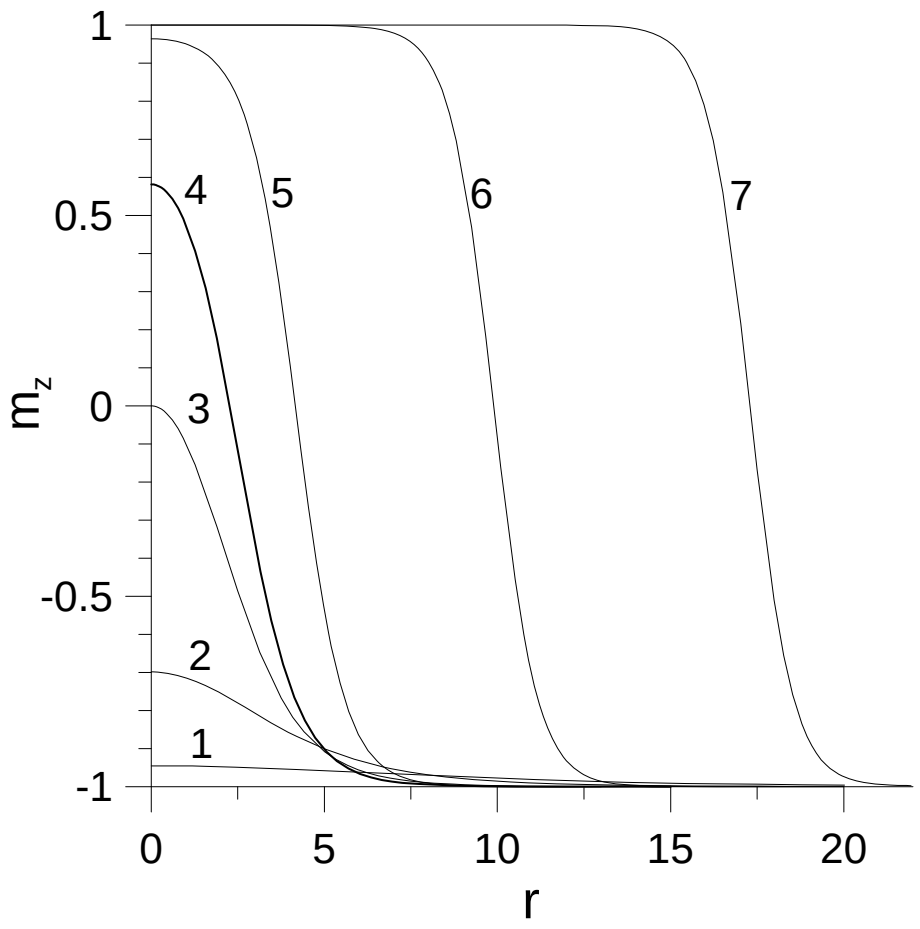

Fig.1 Configurations for the following values of frequency and energy of PBS, if $h = 0.998$:
(1) $\omega = 0$, $E_s = 21.047\,meV$;
(2) $\omega = -0.01$, $E_s = 29.2086\,meV$;
(3) $\omega = -0.053573$, $m_{zm} = 0$; $E_s = 43.605\,meV$;
(4) $\omega = -0.1298$, $E_s = -0.0014\,meV$;
(5) $\omega = -0.3$, $E_s = -991.8\,meV$;
(6) $\omega = -0.5$, $E_s = -29.194\,eV$;
(7) $\omega = -0.57$; $E_s = -188.6\,eV$.

In Fig.2, the frequency dependencies of energy and amplitude for several values of *h* are shown. It is obviously that the PBS are possible in absolutely unstable region too, i.e. at $h > 1$. Here and in all subsequent examples we use the following values: $M_0 = 0.5\times10^{14}\frac{eV}{Oecm^3} \cong 80Oe$ , $K_1 = 1000Oe$, $\alpha = \alpha_z = 3\times10^{-10}Oecm^2$.

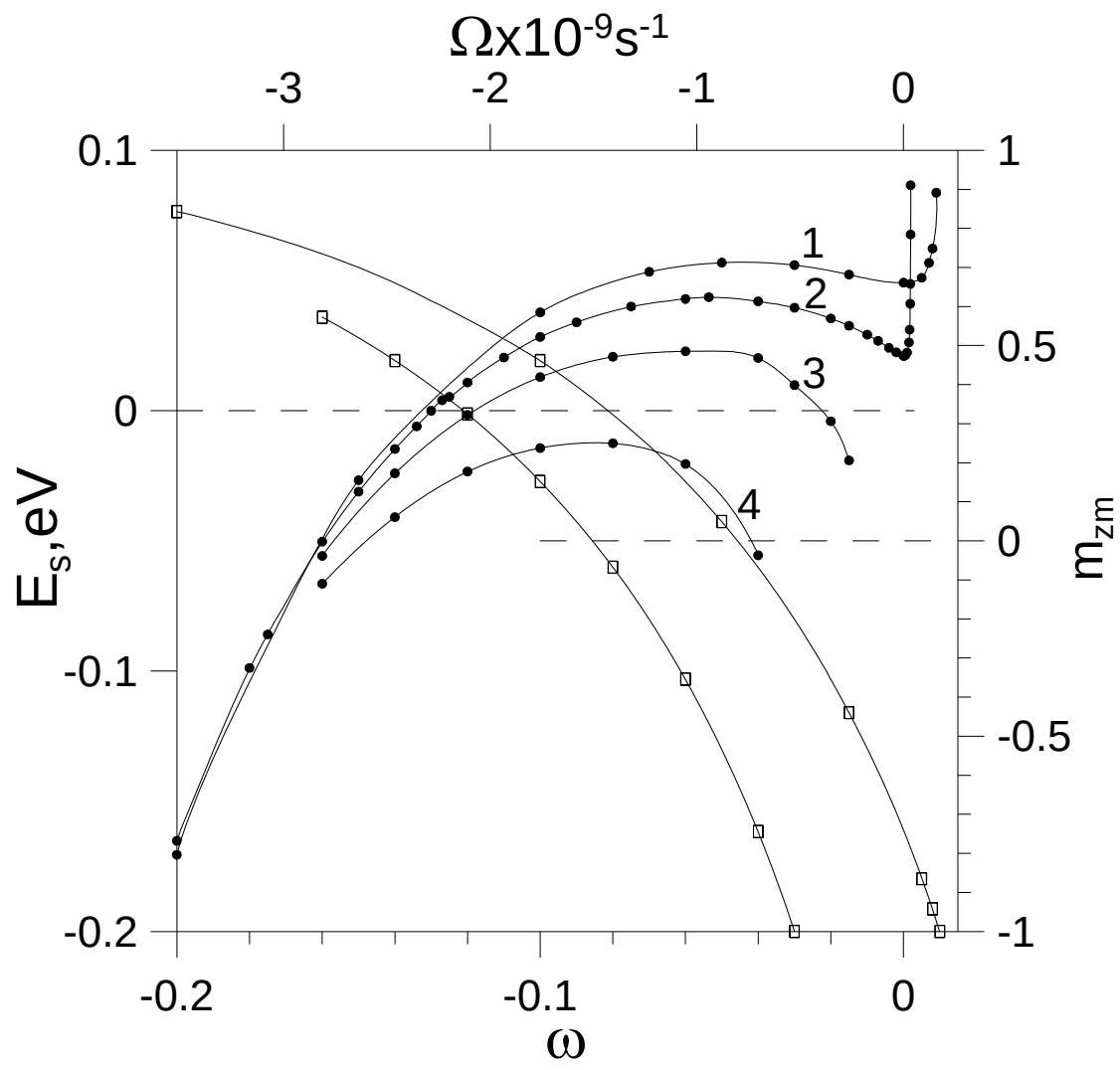

Fig.2 The frequency dependencies of the PBS energy at $k \equiv 0$ for the following values of $h$: (1) 0.99, (2) 0.998, (3) – 1.01, (4) 1.03. The amplitude values $m_{zm}$ are shown only for $h = 0.99$ (upper curve) and for $h = 1.03$ (lower curve). In the upper scale, the frequency in $sec^{-1}$ units has been shown for the same parameters as in Fig. 1. Here and in all subsequent figures, the values of energy are denoted by full circles; the amplitude ($m_{zm}$ or $p_m$), by empty squares; the frequency, by a continuous line; the radius, by crosses.

Comparing two spectra of PBS at $h = 0.998$ in Fig.3, you can see the influence of demagnetizing fields. In the case of curves 2, the parameter $D$ corresponding to $M_0 = 80Oe$ has been used. It is obvious that the $E_s(\omega)$ dependency considerably changes but the main features in the spectrum of PBS remain: the minimum of energy is at $\omega = 0$, the maximum of energy is at $m_{zm} = 0$, energy is negative at sufficiently large frequency (in absolute value), there is a point where the energy changes the sign, i.e. there is the bifurcation point.

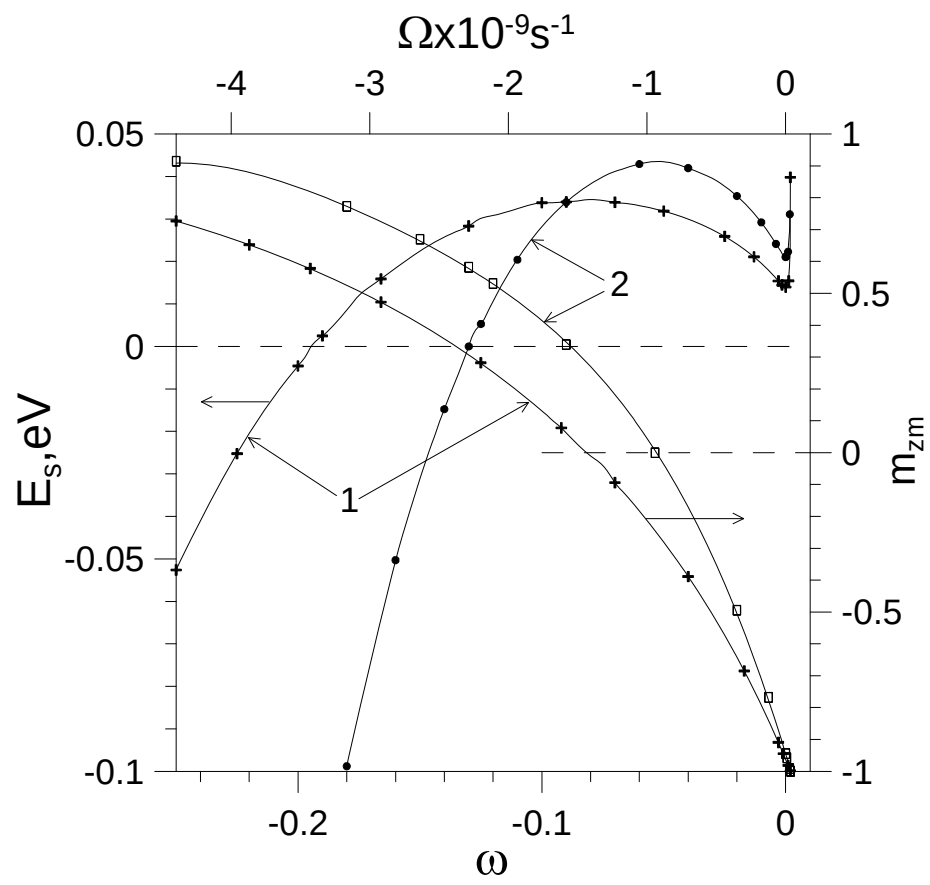

Fig.3 Comparing two spectra of PBS at $h = 0.998$: here the values of energy and amplitude (curves 1) without account of demagnetizing field are denoted by crosses, the other curves (curves 2) correspond to PBS taking into account the demagnetizing fields at $M_0 = 80Oe$ .

**Formatted:** Indent: Left: 0 pt, First line: 14,2 pt, Right: 0 pt, Tabs: Not at 35,45 pt + 453,55 pt

The utmost frequency for PBS corresponds to the frequency of ferromagnetic resonance for metastable state: $\omega_{res1} = (1-h)$. The frequency $\omega_{res2} = -(1+h)$ corresponds to ferromagnetic resonance for stable state of ferromagnet, i.e. at $m_z = +1$. At increasing absolute value of frequency, the amplitude and size of PBS increase. The utmost frequency of the PBS equals $\omega_{lim} = -h$ and, correspondingly, utmost radius of PBS is $r_{1/2} \to \infty$. As you see, the relation $(\omega + h) = 0$ satisfies Eqs. (15) and (20) and in this utmost case $p \to 0, m_z \to +1$.

## 4. Probability of PBS

We divide the processes connected with PBS into two stages, which can be considered in consecutive order. The first stage is a spontaneous birth of soliton, as a result of fluctuations of the energy or of any other processes; the second one is the further evolution of soliton. Such division is fair if we believe that the lifetime of soliton is much more, than the time necessary for its formation [8].

Probability of the spontaneous creation of PBS connected with the fluctuations of energy (and their configuration) at non-zero temperatures is proportional to probability of these fluctuations. Probability of fluctuation in the equilibrium state of a system is expressed by Einstein formula:

$$P_{fl} \approx exp\left(\frac{\Delta S}{k_B}\right), \tag{22}$$

where $\Delta S < 0$ is the entropy change corresponding to the fluctuation. Applying this expression for fluctuations in metastable state of our magnetic system, we can use the following expression for the probability of PBS creation with the $E_S$ energy:

$$P_s = \begin{matrix} A_{1+}\exp\left(\frac{-E_s}{k_B T}\right) \text{if } E_s > 0 \\ A_{1-}\exp\left(\frac{E_s}{k_B T}\right) \text{ if } E_s < 0 \end{matrix}; \tag{23}$$

here $A_{1+}$ and $A_{1-}$ are the configuration coefficients in general depending on ω, $E_S$ and configuration of PBS. We use the (23) expression for PBS at phase transition in ferromagnet, similar to the case of spin-flop transition in antiferromagnet [6]. Considering insignificant quantity of PBS, their small sizes in comparison with the volume of a crystal and slow evolution of solitons, it is possible to suppose, that formula (23) represents the distribution of PBS on the energy in quasi-equilibrium condition existing in an initial stage of phase reorganization.

For the majority of ferromagnetic substances with uni-axial anisotropy, absolute values of the PBS energy, in a wide range of precession frequency, exceed the value $k_BT$ for the temperatures corresponding to the magnetic ordering of crystals. Therefore, the probability of spontaneous origin of such PBS is negligibly small. Only in the area of Fig.3, where $E_S(\omega)$ curves pass through the zero value, the probability of spontaneous origin of PBS can be essential.

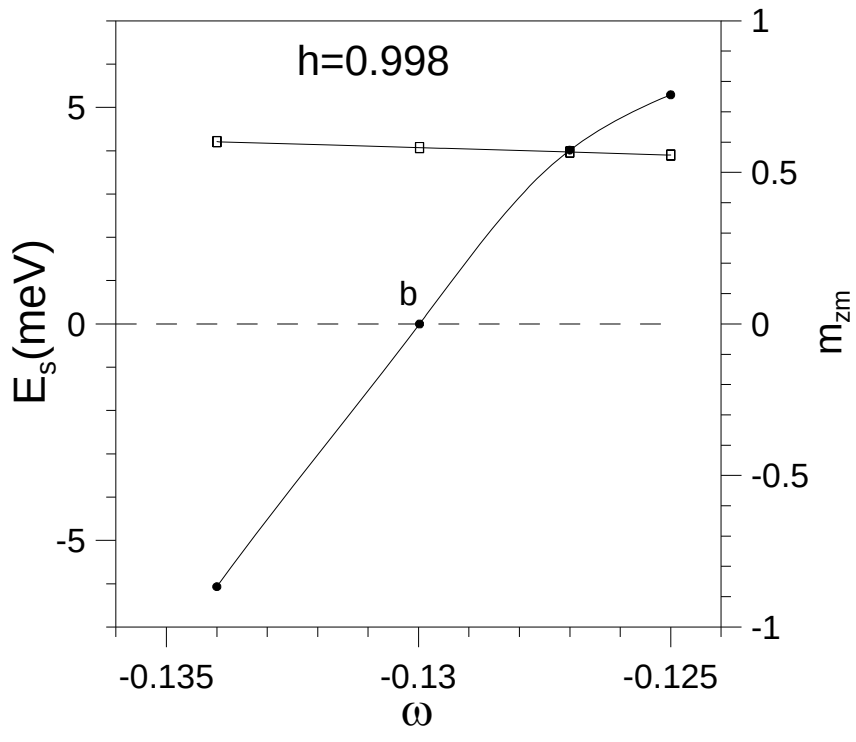

Fig.4 The fragment of Fig. 2 for $h = 0.998$ for $E_s(\omega)$ near the bifurcation point *b*. Frequency of PBS in the *b*-point is $\omega_b \cong -0.13$ , the activation energy of PBS in this point equals zero but the amplitude equals 2 relative to the initial state of the crystal ($m_{z0} = -1$).

In Fig.4 the fragment of Fig.3 for $E_s(\omega)$ dependence around the bifurcation point *b* is presented. Frequency of PBS in the *b*-point is $\omega_b \cong -0.13$ , the activation energy of PBS in this point equals zero.

Exponential dependence of $P_S$ on the PBS energy has been shown in Fig.5. Generally, we have to use differing configuration coefficients for positive and negative energy of PBS near the *b*-point. However, it can be assumed that near the zero value of energy: $A_{1-} \cong A_{1+} = A_1$. Apparently, $A_{1+}, A_{1-} << 1$.

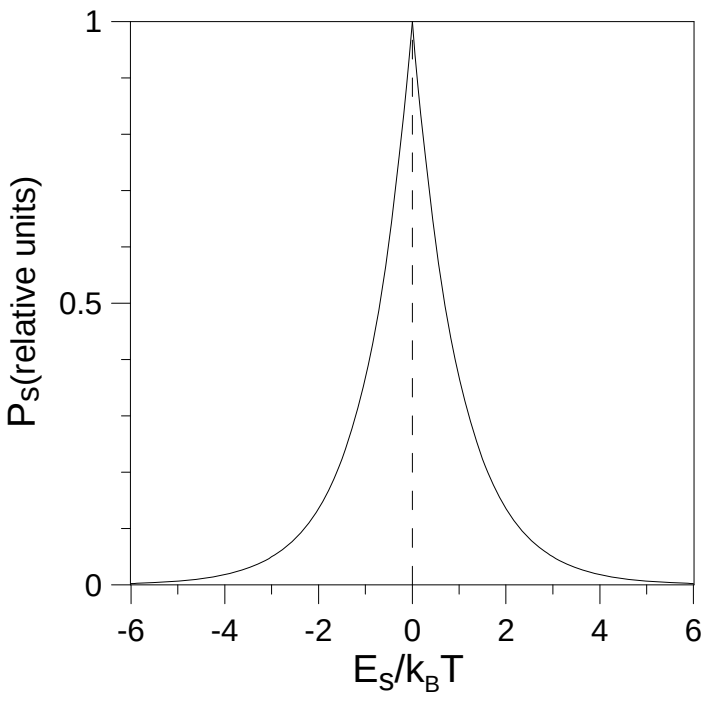

Fig.5 Energy dependence of spontaneous PBS creation probability near the point of bifurcation.

## 5. Time evolution of PBS

To consider the evolution of PBS due the dissipation of energy, it is necessary to take in account that at the dissipation the frequency and pulse of PBS changes in the time, i.e.

$\omega = \omega(\tau)$ and $k = k(\tau)$. Let us consider the behaviour of PBS at $h = 0.998$. Their spectrum is shown in Figs.6 and 7. Energy dissipation of arising PBS changes their amplitude and configuration, and consequently, is accompanied by the change of the frequency, in accordance with (20) equation. In this process the energy of PBS changes continuously along the $E_s(\omega)$ curve.

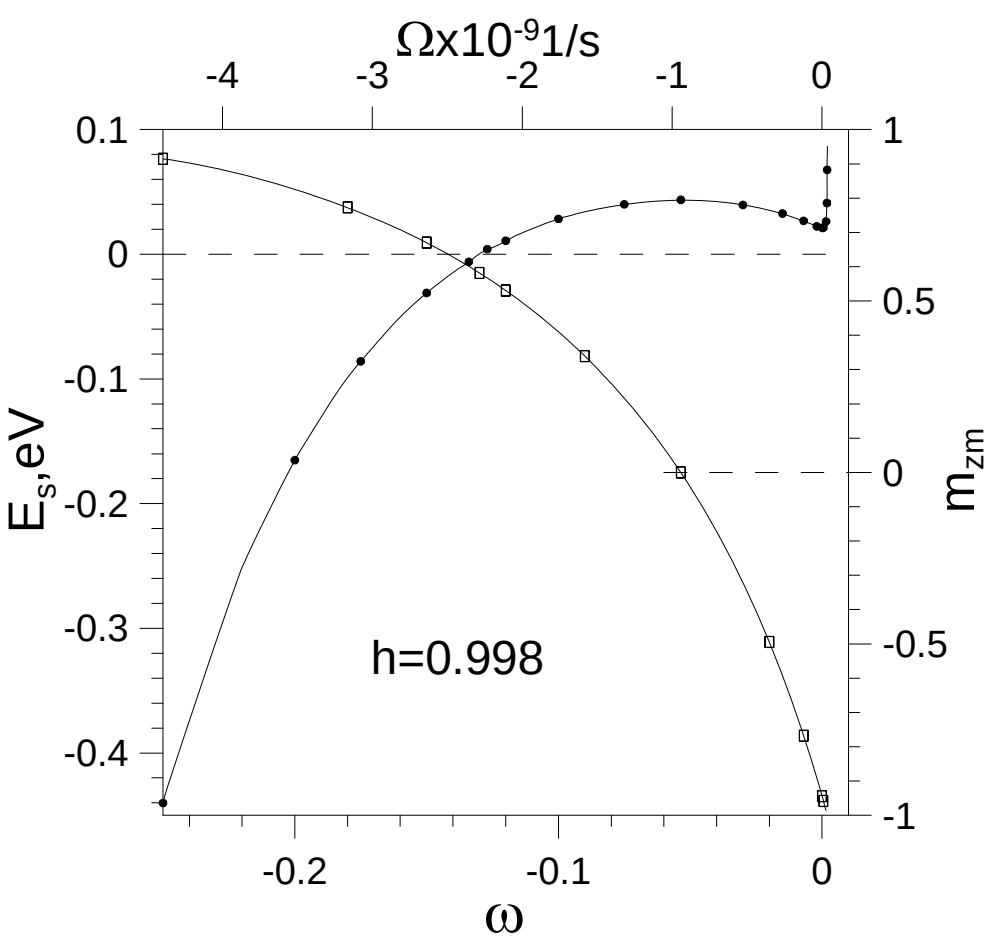

Fig.6 Frequency dependencies of energy and amplitude at $h = 0.998$.

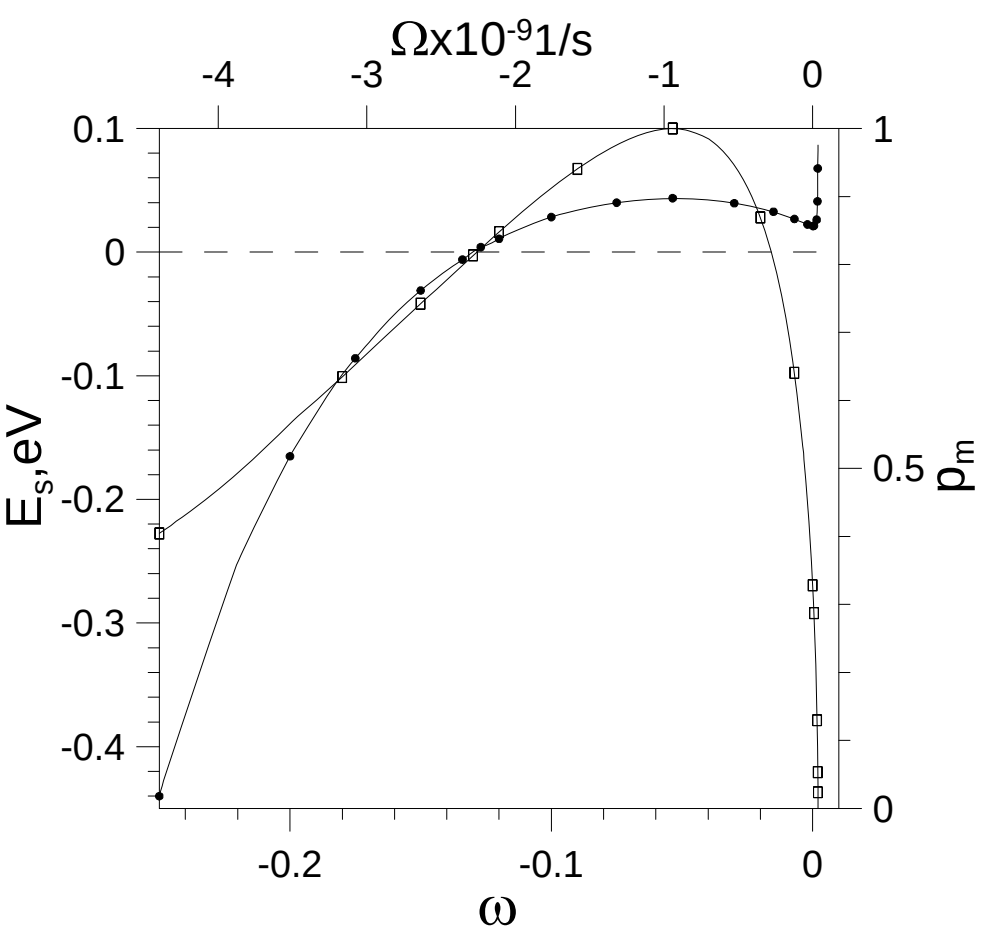

Fig.7 Dependencies of energy and amplitude on the frequency with the same parameters as in Fig. 6, but the amplitude $p_m$ instead of $m_{zm}$ is used.

In Fig.8 the example of the time dependencies of the energy and precession frequency of PBS is presented. Calculations of the PBS evolution have been fulfilled, according to Eqs. (16), (20) and (21), in the following succession:

$$\Delta\omega \to \Delta E_s \text{ (from } E_s(\omega) \text{ dependency)} \to \Delta\tau = -\Delta E_s \Big/ \left(\kappa\omega^2 \int \left(1 - m_z^2\right) r^2 dr\right). \qquad (24)$$

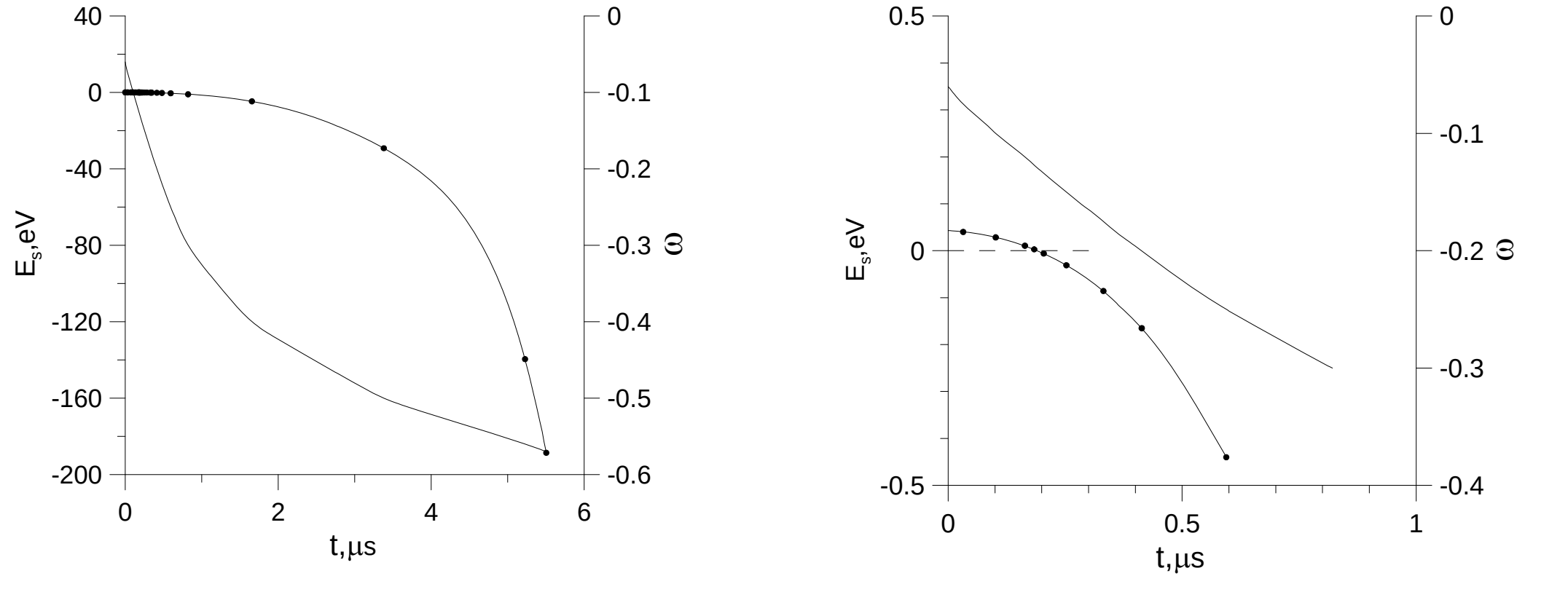

Fig.8 Time dependencies of energy and frequency at $h = 0.998$. In this example the initial values of parameters are $\omega_{init} = -0.06$, $E_{s,init} = 42.9 meV$, $m_{zm,init} = 0.06$ (see the points in Figs. 6 and 7).

Fig.9 The same dependencies as in Fig. 8, but with another scale of energy.

In result, we can see the transformation of PBS into the domain of new phase. The same $E_s(t)$ and $\omega(t)$ dependencies but with another energy scale are shown in Fig.9.

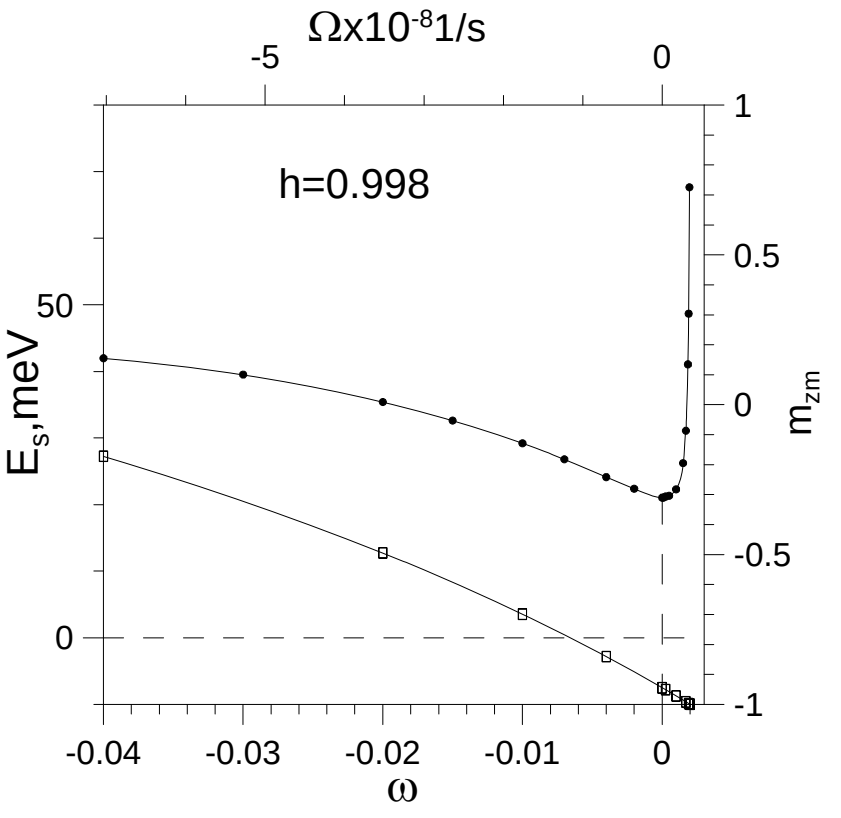

Fig.10 $E_s(\omega)$ and $m_{sm}(\omega)$ dependencies in the region of the minimum of energy for $h = 0.998$.

Another character of PBS can be seen in the range of minimum in $E_s(\omega)$ curve. Such minima exist at $h < 1$, but if $(1-h)$ value is small enough. See, for example, the 1 and 2 curves in Fig.2. In Fig.10, the $E_s(\omega)$ and $m_{sm}(\omega)$ dependencies with minimum are shown for $h = 0.998$ at $\omega > -0.04$. In Figs. 11 and 12, corresponding time dependencies for the energy and frequency are presented using the calculation in accordance with the (24) procedure.

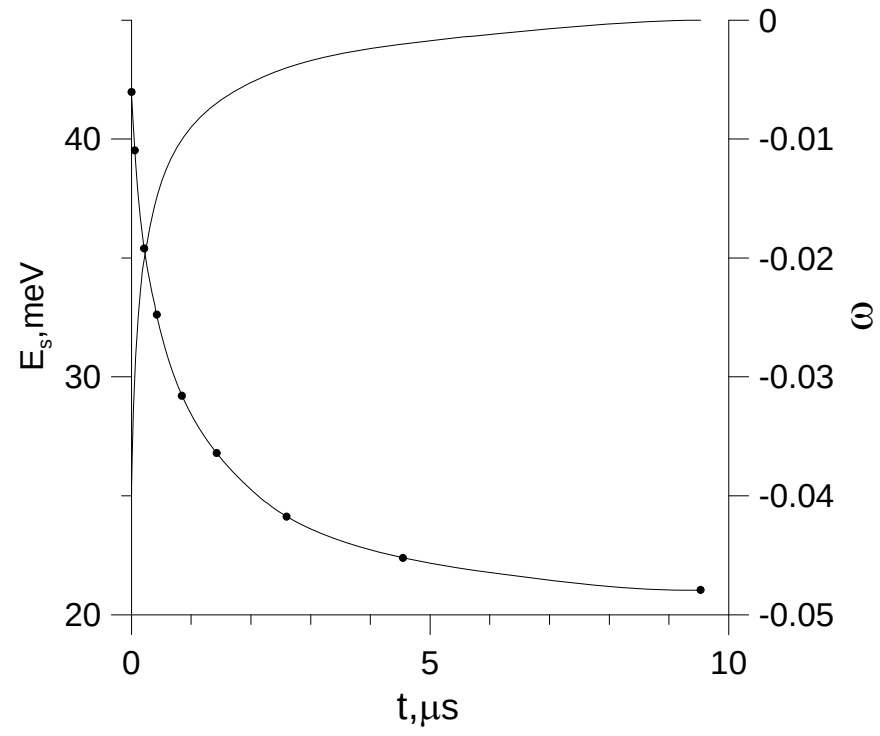

Fig.11 Time dependencies of the energy and frequency for $h = 0.998$ if the initial parameters are $\omega_{init} = -0.04$, $E_{s,init} = 41.9 meV$, $m_{zm,init} = -0.1722$, and end frequency equals zero.

Thus, the PBS arising at small values of $|\omega|$, transform into non-precessing PBS corresponding to the minimum of energy. In the given example, $E_{s,\min} \cong 21.0 meV$. This minimum is a "trap" accumulating PBS. Figure 13 corresponds to Fig. 12, but here the changes of radius and amplitude in the time are shown. It is seen that in process of the evolution, in the given case the radius of PBS decreases thirty times approximately, and correspondingly the amplitude increases considerably.

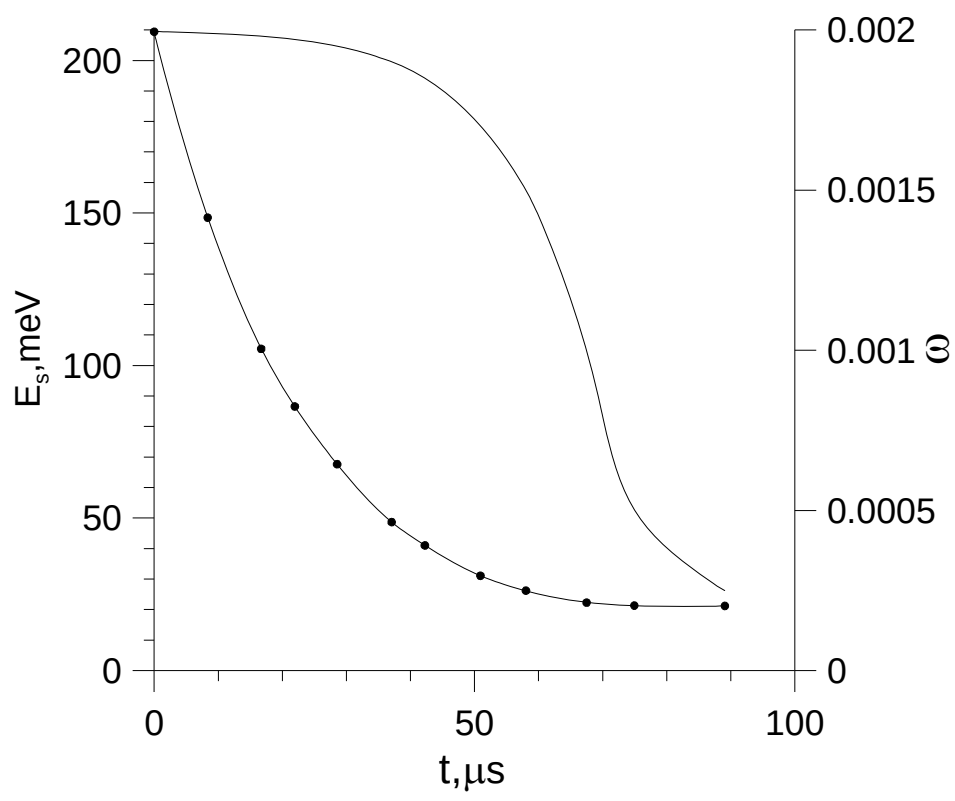

Fig.12 Time dependencies of the energy and frequency for $h = 0.998$ if the initial parameters are $\omega_{init} = +0.001995$, $E_{s,init} = 200.4meV$, $p_{m,init} = 0.0168$, and end frequency equals zero.

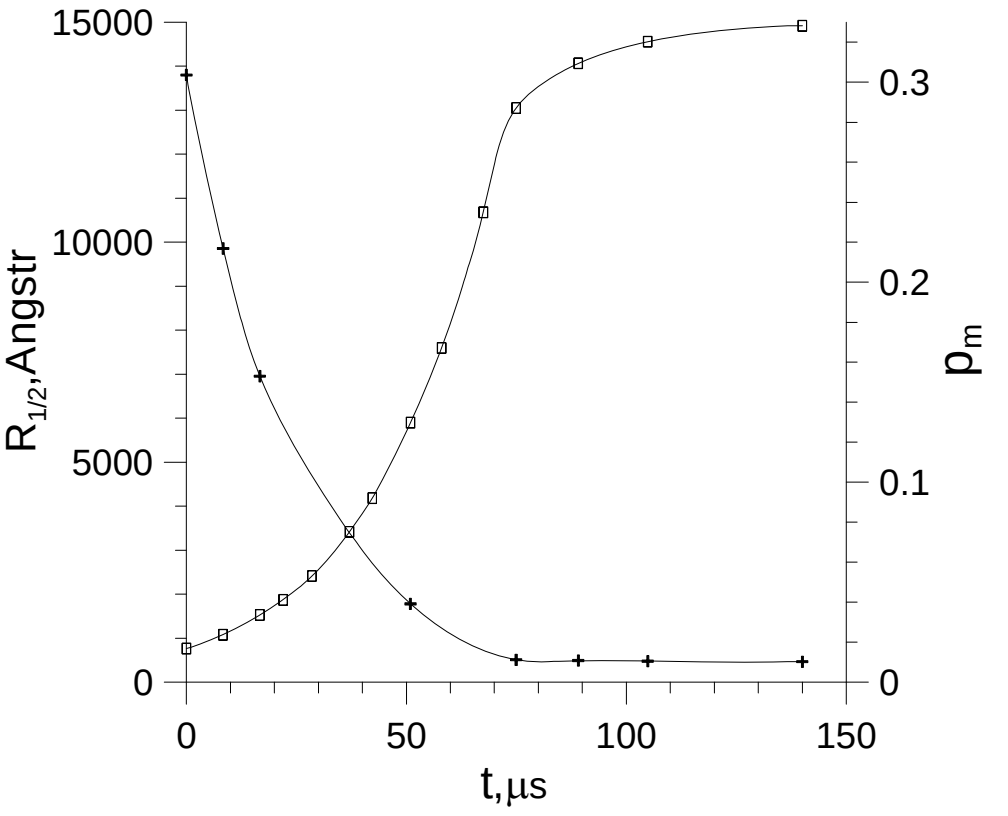

Fig.13 Time dependencies of radius and amplitude of PBS, corresponding to Fig. 12.

Formation of macroscopic domains arising from PBS is possible only at a sufficiently large anisotropy. If the sample of the form that ensures the homogeneity of demagnetizing field $H_{dem0}$ is used and we consider the PBS with the ellipsoid-of-rotation symmetry, the ferromagnets with uni-axial anisotropy can be divided in two groups. If $K_1 > 2N_s M_0$, PBS are possible with amplitude $m_{sm} \to 1$ and negative energy. In such case, PBS mechanism of phase transition exists. But if $K_1 < 2N_s M_0$, such mechanism for phase transition does not exist. In the latter case, the energy of all solitonic states is positive. In Figs. 14 and 15, these two cases are illustrated.

In these Figs, $H_{z,k}$ is the value of external field that is necessary to keep the sample in one-domain state at $m_z = -1$ (assume that we have created such conditions that the sample consists of one domain only). At such external field, the field acting on magnetic moments is $(H_{z,k} + N_0 M_0) < 0$. In the $t_0$ moment, the external field increases so that the acting field $(H_z + N_0 M_0)$ is near to the field of anisotropy $K_1$. For the PBS with $m_{zm} \cong +1$ and large radius, own field of demagnetization is $(-2N_s M_s)$. A necessary condition for existence of such PBS is $(H_z + N_0 M_0 - 2N_s M_s) > 0$ (in our case $M_s = M_0$), i.e., in any case, it is necessary that $2N_s M_0 < K_1$. Two cases are illustrated in Fig.14 and Fig.15 for plane sample with the spherical PBS. The condition $2N_s M_0 < K_1$ is satisfied in the diagram of Fig.14, but is not fulfilled in Fig.15.

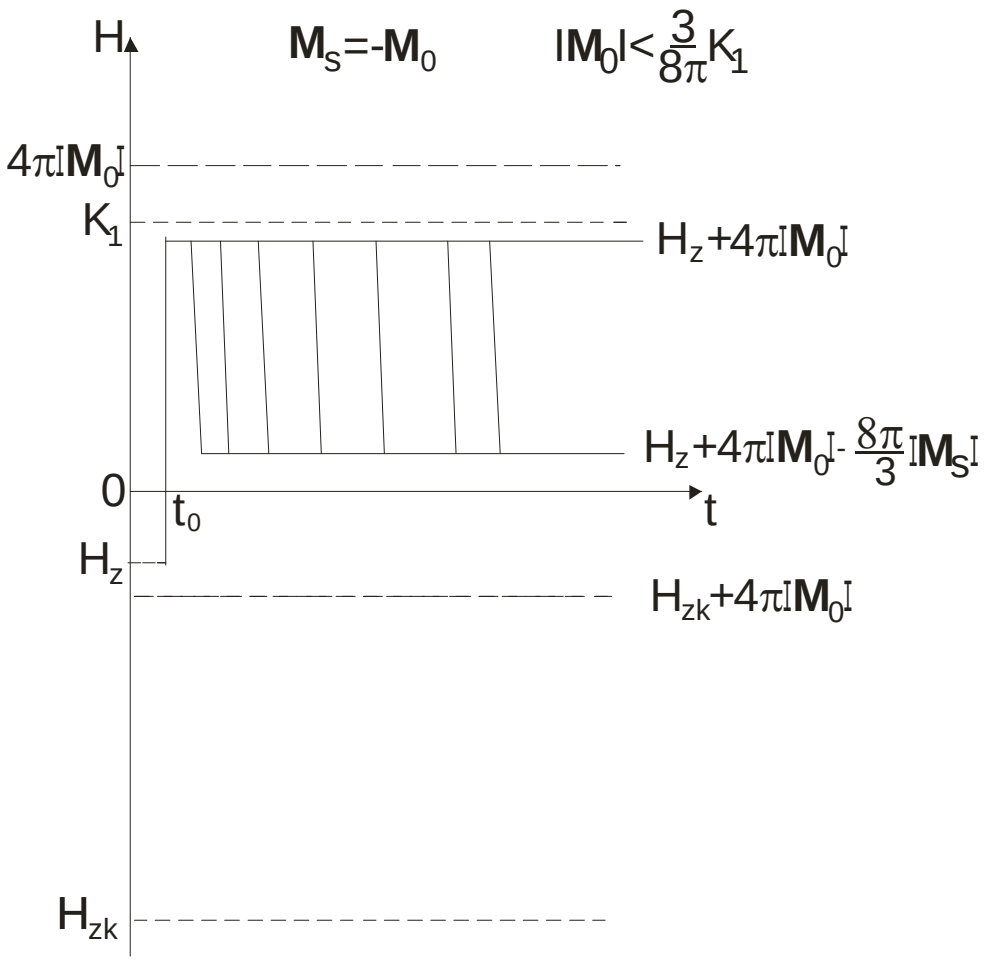

Fig.14 The illustration of relation between the field of anisotropy and demagnetizing field when $K_1 > 2N_z|\mathbf{M}_0|$. In such cases spontaneous origin of PBS with the energy near the bifurcation point and the transformation of them into domains of new phase are possible.

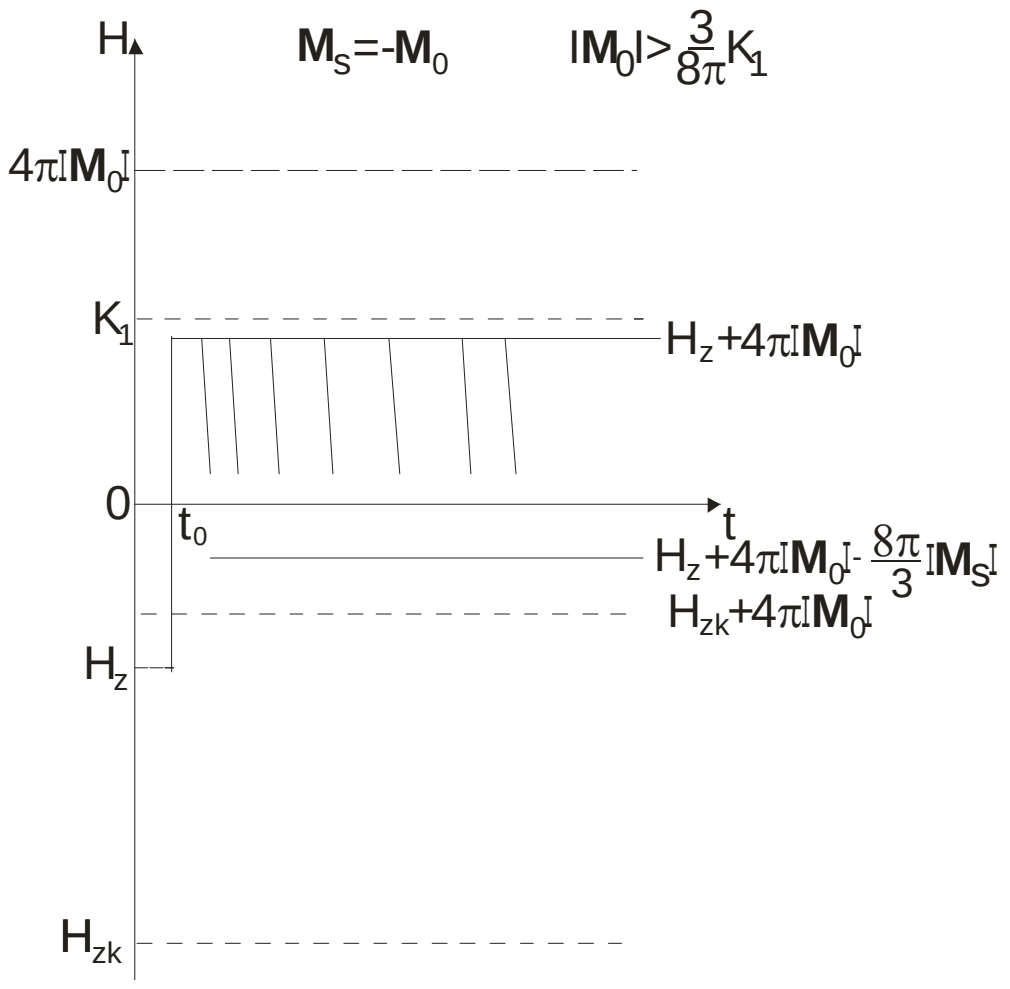

Fig.15 The illustration of relation between the field of anisotropy and demagnetizing field when $K_1 < 2N_z|\mathbf{M}_0|$. In this case the energy of all PBS is positive; and therefore the solitonic mechanism of phase transformation is impossible.

## 6. The change of PBS due to their movement

Now, we consider the influence of movement of PBS on its form. At such condition, the PBS lose their spherical form. From Eq. (12) we have

$$\frac{\partial m_z}{\partial \tau} = -2km_z \frac{\partial m_z}{\partial x} + \kappa(1 - m_z^2)\frac{dk}{d\tau}x - \kappa(1 - m_z^2)\omega(\tau). \tag{25}$$

Let us present the $m_z(x,y,z,\tau)$ function as $m_{zs}(x_s,y,z,\tau)$, where $x_S = (x - v_0\tau)$. In this case

$$-v_0\frac{\partial m_{zs}}{\partial x_s} + \frac{\partial m_{zs}}{\partial \tau} = -2k\frac{\partial m_{zs}}{\partial x_s} + 2k(1 - m_{zs})\frac{\partial m_{zs}}{\partial x_s} + \kappa(1 - m_{zs}^2)\frac{dk}{d\tau}x - \kappa\omega(1 - m_{zs}^2). \tag{26}$$

One can suppose that $v_0(\tau) = 2k$ is the velocity of movement of PBS as a whole. Then, in spherical coordinates, we have the following expression:

$$\frac{\partial m_{zs}}{\partial \tau} = v_0 (1 - m_{zs}) \frac{\partial m_{zs}}{\partial r_s} sin\theta cos\varphi + \kappa \frac{(1 - m_{zs}^2)}{2} \frac{dv_0}{d\tau} x - \kappa (1 - m_{zs}^2) \omega(\tau), \qquad (27)$$

where $r_S$ is the radial coordinate in the system of moving soliton. The first term to the right in Eq. (27) describes the deformation of PBS because of its movement along the *x*-axis. For our PBS $\frac{\partial m_{zs}}{\partial r_s} \cong \frac{\partial m_{zs}}{\partial r} < 0$. Consequently, the frontal side of PBS decreases, i.e. it becomes steeper at the movement than for spherical form, but the back side becomes more declivous to the same extent. The second term corresponds to change of PBS due to the dissipation of energy. For the energy change due to the movement of PBS we have the following:

$$\left( \frac{de(r,\tau)}{d\tau} \right)_{dissk} \cong -\kappa v_0 (\tau)^2 \frac{m_z^2}{1 - m_z^2} \left( \frac{\partial m_z}{\partial x} \right)^2 . \qquad (28)$$

The dissipation of energy leads to the decrease of the velocity, $v_0 \to 0$.

## Conclusions

1. If the ferromagnet with a uni-axial anisotropy has been placed in the magnetic field, antiparallel to the magnetization direction, each value of a magnetic field there corresponds to a continuous spectrum of precessing ball solitons, differing by the frequency and energy.
2. For ferromagnets with relatively large anisotropy, the PBS states include a range of negative values of energy. Probability of spontaneous origin of such PBS increases anomalously when their energy is near zero relative to the energy of the initial phase.
3. The dissipation of energy is accompanied not only by a decrease of the velocity of PBS, but also by a change in precession frequency. Accounting for this change allows us to describe the entire process of continuous transformation of arising PBS.
4. For each value of the magnetic field, corresponding to a metastable state of the ferromagnet, the spectrum of PBS consists of two parts. In the first part, the amplitudes of arising PBS correspond to deviations of ferromagnetism vector from the axis of anisotropy by the angle larger than $90°$. Such PBS have negative precession frequency, and they continuously transform into macroscopic domains of a new phase.
5. In another part of the spectrum, i.e. with less than $90°$ deviations of the initial value of the amplitude from the axis of anisotropy, the precession frequency can be positive or negative, and the minimum of the energy corresponds to $\omega = 0$. In such cases, the energy dissipation is accompanied by a deceleration of precession, i.e. $\omega \to 0$. This minimum at $\omega = 0$ is a "trap" for the accumulation of PBS.

6. In this paper, the time dependencies for the process of transformation of PBS into macroscopic domains of the new phase and for the process of "falling down" of PBS into "trap" are considered.
7. The spectra of PBS, exiting in the overcritical region of the ferromagnet, already in the absence of metastability, i.e. at $h > 1$, have been obtained too. It is assumed that such PBS may be manifested during disintegration of the initial phase.
8. The character of the changes of PBS for the forward movement of the soliton as a whole has been considered. When moving, PBS loses its spherical configuration. The front of soliton, in the direction of motion, becomes steeper, but the back becomes more declivous. At the energy dissipation, the velocity of the soliton decreases and its shape is approaching to spherical configuration.

## Acknowledgement

Author is very grateful to Prof. A.Zvezdin for useful proposal to analyze the solitons at the phase transition in ferromagnet.

**Formatted:** Line spacing: 1.5 lines